\documentclass[twocolumn]{aastex631}
\usepackage{graphicx} 
\usepackage[utf8]{inputenc}
\usepackage{amsmath}
\usepackage{amssymb}
\usepackage{bm} 
\usepackage{gensymb}
\usepackage{csquotes}
\MakeOuterQuote{"}
\usepackage{hyperref}
\usepackage{xcolor}
\usepackage{soul}
\usepackage{multirow}
\usepackage{placeins}
\usepackage{afterpage}

\begin{document}

\title{Identification of Candidate Halos Hosting Massive Black Hole Seeds in the \textit{Renaissance} Simulations with Support Vector Machines}

\author[0000-0003-4811-9863]{Brandon Pries}
\affiliation{Center for Relativistic Astrophysics, School of Physics, Georgia Institute of Technology, \\
837 State Street, Atlanta, GA 30332, USA}

\author[0000-0003-1173-8847]{John H.\ Wise}
\affiliation{Center for Relativistic Astrophysics, School of Physics, Georgia Institute of Technology, \\
837 State Street, Atlanta, GA 30332, USA}

\correspondingauthor{bpries3@gatech.edu}

\date{April 27, 2026}

\begin{abstract}
The nature of the origins of supermassive black holes remains uncertain. Multiple possible seeding pathways have been proposed across a variety of mass scales, each with their own strengths and weaknesses. One such channel is a direct collapse black hole (DCBH), thought to form from the deaths of supermassive stars in pristine atomic cooling halos in the early universe. In this work, we investigate the ability to identify halos likely to form a DCBH based on their properties using a support vector machine (SVM). We implement multiple methods to improve the accuracy of the model, including selecting subsets of critical features and optimizing SVM hyperparameters. We find that our best model requires quantities relevant to star formation, such as the metallicity, incident flux of Lyman-Werner radiation, and halo stellar mass. The SVMs produced from this work can serve as probabilistic and holistic seeding prescriptions for DCBHs in cosmological simulations.
\end{abstract}


\section{Introduction} \label{sec:introduction}

Supermassive black holes (SMBHs) at high redshift remain a topic of intense discussion. Recent observations with telescopes such as the Hubble Space Telescope, James Webb Space Telescope (JWST), and Chandra X-ray Observatory have revealed a plethora of high-redshift SMBHs \citep[e.g.,][]{Bunker2023,Cameron2023,Goulding2023,Larson2023,Abuter2024,Bogdan2024,Bosman2024,Furtak2024,Kovacs2024,Maiolino2024(a),Maiolino2024(c),Maiolino2024(d),Natarajan2024,Scholtz2024,Ji2025(a),Ji2025(b),Napolitano2025(a),Napolitano2025(b),Tripodi2025}. However, how these black holes formed and how they grew to their observed masses remains uncertain.

There are three primary formation mechanisms for the seeds of SMBHs that are delineated by mass and most likely creates a continuous initial mass function for seed black holes \citep{Bromm2003,Volonteri2010(a),Volonteri2012,Smith(Aaron)2017(a),Haemmerle2020(a),Inayoshi2020,Sassano2021,Volonteri2021,Izquierdo-Villalba2024,Regan(John)2024,Yakubu2024,Aggarwal2025(a),Urrutia2025}. The first are "light" seeds, with masses of $M_{\bullet} \lesssim 10^{2} \, \mathrm{M}_{\odot}$, which form from the deaths of the first generation of massive stars (Population~III stars; Pop~III stars) \citep{Alvarez2009,Banik2019,Singh2023,Mehta2024,Cammelli2025(a)}. The light seed formation pathway faces the challenge of needing to grow rapidly via a combination of accretion and mergers; if growth happens solely via accretion, this accretion would need to occur at super-Eddington rates. Although this could be possible for some light seeds \citep{Lupi2016,Pezzulli2016,Lupi2024(a),Lupi2024(b),Mehta2024}, it is unclear if this is common enough for the light seed population as a whole to reproduce the observed SMBH masses in the early universe \citep{Johnson2013,Regan(John)2020(b)}.

The second are intermediate-mass black holes (IMBHs), with masses of $10^{2} \, \mathrm{M}_{\odot} \lesssim M_{\bullet} \lesssim 10^{4} \, \mathrm{M}_{\odot}$, which form from collisions and mergers between stars and light seeds in dense nuclear star clusters \citep{Ryu2016,Sakurai2017,Reinoso2018,Sakurai2019,Regan(John)2020(c),Gaete2024,GonzalezPrieto2024,Vergara2024,Woods2024,Baumgarte2025,Reinoso2025,Vergara2025(b)}. Under a higher star formation efficiency, gas in the halo can be directed into the formation of stars (and subsequently light seeds); the creation of a sufficient number of stars and light seeds in the center of the halo is expected to lead to runaway gravitational collapse via mergers, collisions, and tidal disruption events \citep{Rantala2025(b)}. Inflowing gas could lead to increased interaction rates due to dynamical friction \citep{Reinoso2018}, but the interaction processes themselves, along with stellar feedback, could serve to disrupt the gas supply to the remnant IMBH, limiting opportunities for future accretion into an SMBH \citep{Gaburov2010}. It is also believed that IMBHs are unlikely to form if the metallicity is sufficiently high $\left(Z \gtrsim 10^{-3} \, \mathrm{Z}_{\odot}\right)$ as this leads to larger mass losses during stellar collisions \citep{Gaburov2010} and leads to more widespread star formation throughout the halo, limiting the gas supply to the central cluster \citep{Devecchi2009}.

The third are "heavy" seeds, commonly referred to as direct collapse black holes (DCBHs), with masses of $M_{\bullet} \gtrsim 10^{4} \, \mathrm{M}_{\odot}$, which form from the collapsed cores of supermassive stars (SMSs) \citep{Latif2013,Choi2015,Agarwal2016(a),Suazo2019,Chon2020,Prole2024(a)}; we will focus on this formation channel for the remainder of this paper. DCBHs are believed to form in pregalactic halos that reach the atomic cooling limit (ACL) that roughly corresponds to $M_{\mathrm{halo}} \sim 10^{8} \, \mathrm{M}_{\odot}$ and $T_{\mathrm{vir}} \sim 10^{4} \, \mathrm{K}$ at $z \approx 15$ \citep{Loeb1994,Shang2010}. At this limit, the halo is able to efficiently cool via atomic emission and the gas within the halo begins to collapse. Under normal circumstances, the presence of coolants such as metals and molecular hydrogen ($\mathrm{H}_{2}$) leads to strong fragmentation of the gas and the formation of stars and star clusters; this would likely produce light or intermediate-mass seeds. However, with sufficient absence of these coolants, fragmentation can be inhibited, leading to a buildup of material in the center of the halo. This is expected to lead to an SMS with a mass of $M_{\mathrm{SMS}} \sim 10^{4} - 10^{5} \, \mathrm{M}_{\odot}$. If fragmentation occurs on small scales, this could instead produce a dense star cluster of massive stars with a comparable total mass; however, collisions and mergers between these stars are expected to lead to the eventual formation of an SMS \citep{PortegiesZwart2004,Regan(John)2020(c)}. At these mass scales, the core of an SMS becomes unstable under general relativity and collapses into a black hole of mass $M_{\bullet} \sim 10-20 \, \mathrm{M}_{\odot}$. The core collapse can be further catalyzed by neutrino cooling via the Urca process \citep{Begelman2006,Zwick2023}. Once this black hole forms, it can efficiently accrete a large fraction of the surrounding envelope of the SMS in just a few million years \citep{Begelman2008,Volonteri2010(b),Coughlin2024,Fujibayashi2025}. The heavy seed formation mechanism circumvents the requirement for super-Eddington accretion necessary for light seeds to grow to supermassive scales in a few hundred million years. Although potential high-$z$ candidates for DCBHs have been observed \citep[e.g.,][]{Agarwal2016(b),Pacucci2016,Smith(Aaron)2016,Nabizadeh2024,Natarajan2024,Juodzbalis2025(b)}, further observations are required to try to determine the seeding mechanisms of these SMBHs.

For the heavy seed pathway to be viable, there must be some mechanism to suppress or destroy coolants within the halo. The primary source of metals in the early universe is star formation, so the halo must have no prior star formation. The halo must also not be chemically enriched above a critical metallicity threshold $\left(Z \sim 10^{-3} \, \mathrm{Z}_{\odot}\right)$ from its own SMS(s) or from stars in neighboring halos \citep{Chon2020,Chon2025}. In the absence of metals, $\mathrm{H}_{2}$ is the dominant coolant, which can be photodissociated with sufficient illumination from a nearby source of Lyman-Werner (LW) radiation, ultraviolet (UV) light in the energy band $11.2 - 13.6 \, \mathrm{eV}$ \citep{Oh2002,Dijkstra2008,Wolcott-Green2011(a),Visbal2014(b)}. ACL halos are capable of efficiently cooling via atomic hydrogen down to $8000 \, \mathrm{K}$. If not suppressed, cooling via $\mathrm{H}_{2}$ is efficient down to temperatures of $300 \, \mathrm{K}$, though this process leads to increased fragmentation and the formation of typically massive stars \citep{Bromm2003,Yoshida2003}. With a sufficient flux of LW radiation, this cooling mechanism can be suppressed. The Universe is optically thin in the LW band \citep{Haiman1997(a),Haiman1997(b),Holzbauer2012}, so high-$z$ galaxies can build up a LW flux that could sufficiently suppress $\mathrm{H}_{2}$ concentrations in more pristine halos. Starlight from nearby galaxies can dominate over the LW background, leading to the development of the "close pair" scenario for DCBH formation \citep{Dijkstra2008,Visbal2014(b),Regan(John)2017}. However, the collapse itself serves to increase the $\mathrm{H}_{2}$ concentration within the core, leading to self-shielding from the incident LW radiation \citep{Wolcott-Green2011(a),Regan(John)2014,Wolcott-Green2019,Patrick2023}. At this stage, the core cools via $\mathrm{H}_{2}$ to $\sim\!500 \, \mathrm{K}$, accelerating the collapse and leading to small-scale fragmentation but below the levels of present-day star formation. It is also possible that other mechanisms could serve to destroy $\mathrm{H}_{2}$, such as dynamical heating from rapid halo growth \citep{Wise2019(a),Mayer2024}; strong shocks from cold accretion flows \citep{Inayoshi2012,Fernandez2014,Kiyuna2024}, from supernovae \citep{Mayer2024}, and from protogalaxy mergers \citep{Mayer2010,Mayer2015,Mayer2019,Inayoshi2015(b)}; and the detachment of $\mathrm{H}^{-}$ via infrared (IR) radiation \citep{Wolcott-Green2012,Wolcott-Green2017,Wolcott-Green2021}. Any of these mechanisms would lower the importance of assuming a sufficient flux of LW radiation \citep{Mayer2024}.

The formation of DCBHs is expected to be a complex, nonlinear process dependent on several variables. Previous work by \citet{Mone2025} in determining the importance of physical characteristics relevant to DCBH formation lends itself well to explorations with machine learning (ML). In particular, the machine learning architecture of support vector machines \citep[SVMs;][]{Cortes1995} are well-suited to high-dimensional problems, making them a strong candidate for use in this context. SVMs have been leveraged in other astrophysical work, such as classification of stellar spectra \citep{Zhong-Bao2017} and identifying halos that are expected to host Pop~III star formation \citep{Grace2020}.

The remainder of this paper is structured as follows: we present the methods in Section~\ref{sec:methods}, including a discussion of our dataset in Section~\ref{sec:data} and the implementation of the SVMs in Section~\ref{sec:svm}; we present our results in Section~\ref{sec:results}; we discuss the implications of our results in Section~\ref{sec:discussion}; and we conclude in Section~\ref{sec:conclusion}.

\section{Methods} \label{sec:methods}

\subsection{Data} \label{sec:data}

The \textit{Renaissance} simulations \citep{OShea2015,Xu(Hao)2016} were run with \texttt{\textsc{Enzo}} \citep{Bryan2014,Brummel-Smith2019}, an adaptive mesh refinement (AMR) code for hydrodynamic simulations, specializing in cosmological simulations. The \textit{Renaissance} simulations model ionizing radiation transport with ray tracing \citep{Wise2011}. They used the following cosmological parameters from \textit{WMAP}7 \citep{Komatsu2011}: $\Omega_{m} = 0.266$, $\Omega_{b} = 0.0449$, $\Omega_{\Lambda} = 0.734$, $h = 0.71$, $\sigma_{8} = 0.81$, and $n = 0.963$ \citep{Xu(Hao)2013}. The initial conditions were generated with MUSIC \citep{Hahn2011}. We use the \texttt{yt} analysis tool from \citet{Turk2011}, as well as \texttt{ytree} \citep{Smith(Britton)2019} to analyze halo merger trees.

We investigate several quantities of interest that are potentially relevant to the formation of a DCBH. These include:

\begin{enumerate}
    \item \textit{central halo properties} (averaged within the central $50 \, \mathrm{pc}$), such as the density, temperature, radial velocity, tangential velocity, turbulent (root-mean-square, or RMS) velocity, $\mathrm{H}_{2}$ fraction, LW flux, and radial mass flux;
    \item \textit{halo properties}, such as the (stellar) mass, metallicity, growth rate, and gas and dark matter (DM) spin parameters; and
    \item \textit{environmental properties}, such as the large-scale overdensity, distance to the closest galaxy, and $t_{1}$ tidal field eigenvalue.
\end{enumerate}

For data distributions of these quantities, see Figure~1 from \citet{Mone2025}; for correlations between these quantities, see the right panel of Figure~2 from the same publication.

We consider such quantities as features in our SVM. Some of these quantities are directly tracked during the simulation (such as the temperature), while others are calculated during analysis of the simulation data (such as the halo mass and growth rate, which requires halo finding and handling of halo merger trees). However, some of these quantities may not be intuitive, so we discuss below how such quantities are calculated.

We use the \texttt{\textsc{rockstar}} halo finder \citep{Behroozi2013(a)} to identify the halos and the \texttt{\textsc{consistent-trees}} algorithm \citep{Behroozi2013(b)} to construct the halo merger trees. We adopt the same calculation of the halo growth rate as presented in \citet{Mone2025}. Because \texttt{\textsc{rockstar}} only tracks the dark matter within a halo, we implement a mass correction factor of $\Omega_{m}/(\Omega_{m} - \Omega_{b})$ to account for the baryonic mass.

We calculate the LW flux due to its importance in dissociating $\mathrm{H}_{2}$, which is expected to suppress fragmentation. The LW flux is given by

\begin{equation}
F_{\mathrm{LW}} = \frac{E_{\mathrm{LW}} k_{\mathrm{diss}}}{4\pi^{2} \sigma_{\mathrm{H2}} \nu_{\mathrm{H}}},
\end{equation}

\noindent where $E_{\mathrm{LW}} = 12.4 \, \mathrm{eV}$ is the central energy in the LW band, $k_{\mathrm{diss}}$ is the $\mathrm{H}_{2}$ dissociation rate taken directly from the simulation outputs, $\sigma_{\mathrm{H2}} = 3.71 \times 10^{-18} \, \mathrm{{cm}^{2}}$ is the average effective $\mathrm{H}_{2}$ dissociation cross section in the LW band \citep{Wise2011}, and $\nu_{\mathrm{H}} = 3.29 \times 10^{15} \, \mathrm{Hz}$ is the Rydberg constant for hydrogen.

We also consider the radial gas mass flux because high inflow rates can promote large concentrations of matter in a single location, leading to the development of an SMS and subsequently a DCBH. The spherically averaged radial gas mass flux is given by

\begin{equation}
\dot{m} = -4\pi r^{2} \rho v_{r},
\end{equation}

\noindent where $r$ is the radius, $\rho$ is the density, and $v_{r}$ is the radial velocity.

The gas and dark matter spin parameters can have an impact on DCBH formation because halos with higher spin parameters could be rotationally supported, making it more difficult to drive matter to the center. For the spin parameters, we follow the prescription in \citet{Bullock2001} and calculate them as

\begin{equation}
\lambda = \frac{J}{\sqrt{2} M V R},
\end{equation}

\noindent where $J$ is the angular momentum, $M$ is the mass, $R$ is the radius, and $V$ is the circular velocity given by $V^{2} = GM/R$.

We also consider the tidal eigenvalues, which provide information about the large-scale ($\gtrsim 10 \, \mathrm{kpc}$) matter distribution. We adopt the tidal field equations from \citet{Dalal2008} and \citet{DiMatteo2017}, where the tidal eigenvalues are given by

\begin{equation}
T_{ij} = S_{ij} - \frac{1}{3} \sum_{i} {S_{ii}},
\end{equation}

\noindent where $S_{ij}$ is the strain tensor, given by

\begin{equation}
S_{ij} = \nabla_{i} \nabla_{j} \phi,
\end{equation}

\noindent where $\phi$ is the gravitational potential. Rather than calculate $S_{ij}$ directly, we calculate it in Fourier space as

\begin{equation}
\hat{S}_{ij} = \frac{k_{i} k_{j}}{k^{2}} \, \hat{\delta},
\end{equation}

\noindent where $k$ is the wavenumber and $\hat{\delta}$ is the density in Fourier space. Specified eigenvalues are denoted as: $t_{1}$, the largest positive value; $t_{3}$, the largest negative value; and $t_{2}$, defined by

\begin{equation}
t_{1} + t_{2} + t_{3} = 0.
\end{equation}

\noindent A large $t_{1}$ value indicates a bulk convergence of matter along the direction of the corresponding eigenvector, which is expected to lead to a large-scale overdensity that will contain a massive cluster with massive galaxies at later times. Such a region is more likely to contain a massive, rapidly-growing halo, which is conducive to DCBH formation.

We use the same dataset presented in \citet{Mone2025}, including the same candidacy definition, with two changes. First, we restrict the mass range of the halos to those that are candidates for hosting DCBHs; this range is given by $10^{7.5} \, \mathrm{M}_{\odot} \lesssim M_{\text{halo}} \lesssim 10^{8} \, \mathrm{M}_{\odot}$, where the lower bound roughly corresponds to the ACL at $z \approx 15$ and the upper bound accounts for the time delay between reaching the ACL and the collapse of the halo as it continues to gain mass. We place this restriction because we expect halos with masses outside this region to not be candidates for hosting DCBH formation. Halos that are undermassive would likely have an insufficient gravitational potential to promote rapid infall, and halos that are overmassive would likely have already collapsed to form stars and be chemically enriched. By eliminating these halos from our sample, we restrict the SVM to the appropriate mass range for candidacy. Second, we do not implement the cut that removes halos with metal-enriched stars younger than $20 \, \mathrm{Myr}$. Since the goal of this work is to develop a model that can be used as a seeding prescription, implementing this cut would artificially bias the model by altering the LW distribution from the raw simulation output. With these changes, our dataset consists of $\sim\!8400$ halos, of which $35$ are considered prime candidates for hosting DCBHs. For our machine learning models, the dataset is divided using the same 90\%\slash 10\% train\slash test split for each model.

\subsection{SVM and Stats} \label{sec:svm}

We implement a machine learning model called an SVM \citep{Cortes1995}. An SVM tunes the coefficients of a function to achieve the highest prediction accuracy across a set of data. This fit is subject to hyperparameters, such as the functional form, which we discuss in Sec.~\ref{sec:hyperparameter_tuning}. Graphically, an SVM used for classification (called a support vector classifier, or SVC) is attempting to define a boundary that best separates the datapoints from two or more classes in phase space, and the coefficients of the function describe how strongly this boundary depends on a given variable. SVMs are designed to scale well to high-dimensional data; this makes them well-suited to our use case, where each physical variable corresponds to a new dimension.

We use the SVM software provided by \textit{Scikit-learn} \citep[\texttt{sklearn};][]{Pedregosa2011} using the \texttt{sklearn.svm.SVC} function. We consider three different binary classification scoring metrics: the precision, recall, and $F_{1}$ score, defined as

\begin{subequations} \label{eq:performance_metrics}
\begin{equation} \label{eq:precision}
P = \frac{\mathrm{TP}}{\mathrm{TP} + \mathrm{FP}}
\end{equation}
\begin{equation} \label{eq:recall}
R = \frac{\mathrm{TP}}{\mathrm{TP} + \mathrm{FN}}
\end{equation}
\begin{equation} \label{eq:f1_score}
F_{1} = \left(\frac{\frac{1}{P} + \frac{1}{R}}{2}\right)^{-1} = \frac{2\mathrm{TP}}{2\mathrm{TP} + \mathrm{FP} + \mathrm{FN}}
\end{equation}
\end{subequations}

\noindent where $\mathrm{TP}$ is the number of true positives, $\mathrm{FP}$ is the number of false positives, and $\mathrm{FN}$ is the number of false negatives. Intuitively, the precision describes the ability of the classifier to correctly distinguish between the positive class and negative class (no false positives); the recall describes the ability of the classifier to correctly recover all points belonging to the positive class (no false negatives); and the $F_{1}$ score is the harmonic mean of the two, placing equal weight on both. In the context of our work, the precision describes the ability of the classifier to correctly classify all non-candidates as non-candidates, and the recall describes the ability of the classifier to correctly classify all candidates as candidates. Note that all three achieve a minimum of $0$ and a maximum of $1$.

In order to improve the model beyond standard training, we implement two methods. The first is hyperparameter tuning, which is used to find the optimal set of hyperparameters. Hyperparameters are parameters that affect how the model fits to the training data, but are distinct from the training data. These hyperparameters are typically parameters in the machine learning model's loss function, which determine how heavily the model is penalized for incorrect predictions. The second is feature selection, which is used to find one or more optimal subsets of features. It is possible that extraneous or highly-correlated features could be overburdening the SVM, and removing these features could lead to improved performance. However, these two methods are intertwined: the set of features influences the optimal set of hyperparameters, and the set of hyperparameters influences the optimal set of features. To break this degeneracy, we apply these methods in the following steps:

\begin{enumerate}
    \item[1.] We apply hyperparameter tuning first on the SVM that uses all features.
    \item[2.] Using this optimal set of hyperparameters, we use feature importance ranking strategies to select one or more feature subsets to pursue. By performing hyperparameter tuning first, we have the best model that uses all features and can therefore get the best relative importance measurements among all the features.
    \item[3.] We re-tune the hyperparameters for each individual feature subset chosen in Step~2.
\end{enumerate}

\subsubsection{Hyperparameter Tuning} \label{sec:hyperparameter_tuning}

For hyperparameter tuning, we investigate three key hyperparameters. The first is the kernel $K$, which describes the functional form of the decision boundary in phase space. \texttt{sklearn} supports linear, polynomial, radial basis function (RBF), and sigmoid kernels; we consider all of these, sampling integer polynomial orders in the range $2-5$. The second is the regularization parameter $C$, which determines how heavily to penalize the SVM for incorrect classifications; in terms of the decision boundaries, $C$ also inversely influences the width of the margins on either side of the decision boundary, which determines how many support vectors the SVM uses to fine-tune the shape of the decision boundary. We sample values of $C$ in the range $10^{-2}-10^{4}$ in steps of $0.25 \, \mathrm{dex}$. The third is the class weight $w$, which determines how heavily to weight data points from one class relative to another. This is particularly important for unbalanced problems like ours where only a small fraction of the data belong to a particular class. In our case, $w$ represents how much more weight is applied to candidate halos than non-candidate halos. We sample values of $w$ in the range $10^{0}-10^{4}$ in steps of $0.25 \, \mathrm{dex}$. We search over all possible combinations of these parameters using the \texttt{sklearn.model\_selection.GridSearchCV} function.

\subsubsection{Feature Importance} \label{sec:feature_importance}

\begin{figure*}[!ht]
\centering
\includegraphics[width=0.95\textwidth]{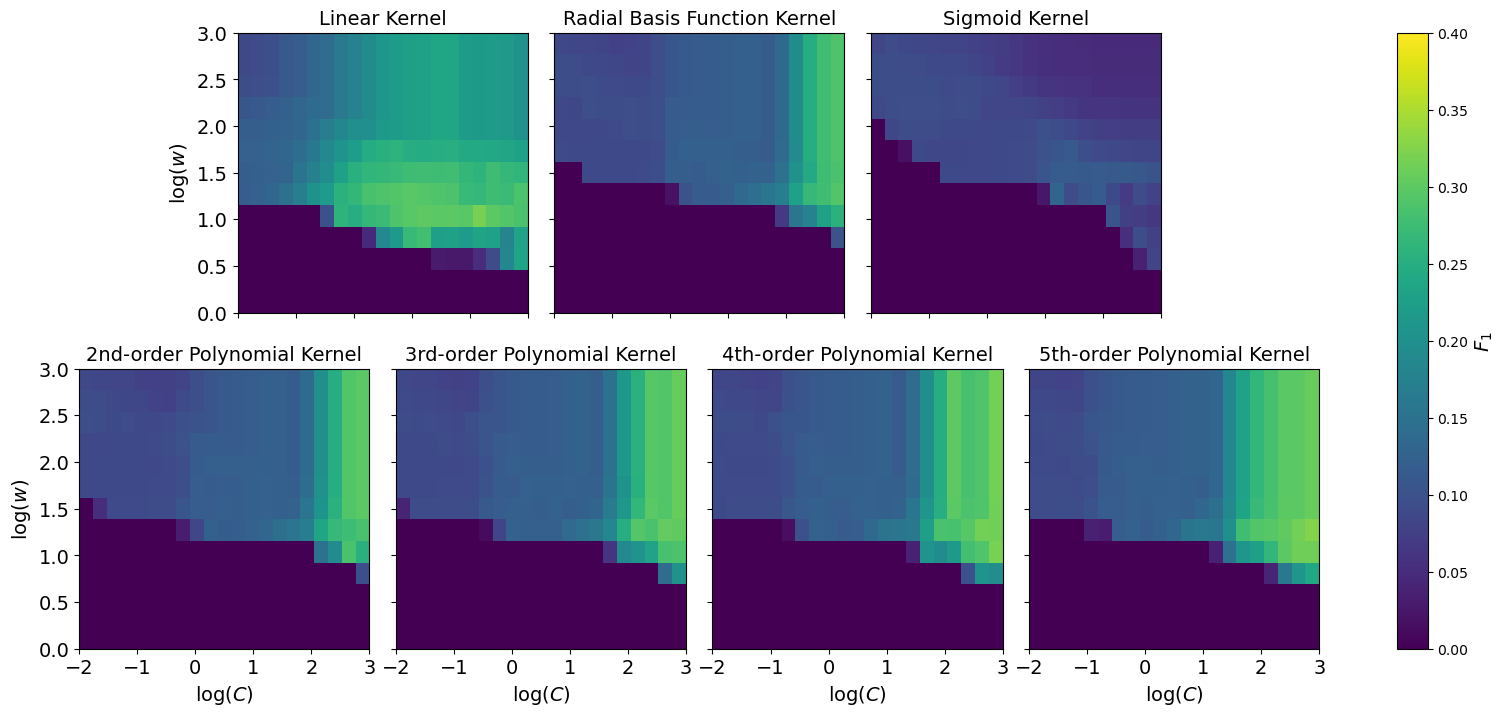}
\caption{\label{fig:F1s_all_features}$F_{1}$ scores for each combination of the regularization parameter $C$ (x-axes) and the class weight $w$ (y-axes) for each kernel tested (panels). The top row shows the linear, RBF, and sigmoid kernels, respectively, and the bottom row shows polynomial kernels of different polynomial orders (2, 3, 4, and 5, respectively). Most kernels show a preference for large values of $C$ and intermediate to high values of $w$.}
\end{figure*}

We consider four different methods for measuring the relative feature importances. The first is SelectKBest, a native \texttt{sklearn} function that selects the $k$ features with the highest importance according to a scoring function. The second is recursive feature elimination (RFE). With RFE, we train a set of models that each has one feature removed and consider the $F_{1}$ scores produced by each of these models. We then determine the model with the highest score, as its high score comes from removing a low-importance feature. We then eliminate this feature from the set of available features, and repeat this process until only one feature remains. This process can also be done in reverse (eliminating the model with the lowest score, corresponding to the feature with the highest importance), yielding results for both the "forward" and "backward" directions, respectively.

The third uses the Mahalanobis distance \citep[see, e.g.,][]{DeMaesschalck2000}. The Mahalanobis distance can be thought of as a generalization of the $Z$ score to multiple dimensions, accounting for correlations between those dimensions. Like with RFE, we calculate the Mahalanobis distances with each feature removed, then eliminate the one corresponding to the lowest change in the Mahalanobis distance as this indicates low importance. Also as with RFE, this can be done in both the "forward" and "backward" directions.

We use the following prescription to calculate the Mahalanobis distance:

\begin{enumerate}
    \item[1.] First, we use the standard definition of the Mahalanobis distance,
    \begin{equation}
    d_{\mathrm{M}} = \sqrt{\left(\mathbf{x} - \bm{\mu}\right)^{T} \sigma^{-1} \left(\mathbf{x} - \bm{\mu}\right)},
    \end{equation}
    \noindent where $\mathbf{x}$ is the vector of values for a particular halo in our phase space, $\bm{\mu}$ is the vector of mean values for the corresponding variables, and $\sigma$ is the covariance matrix. This calculation can fail if $\sigma$ is not invertible.
    \item[2.] If the first method fails, we use an alternative calculation for the Mahalanobis distance that does not require inverting the covariance matrix:
    \begin{equation}
    d_{\mathrm{M}} = \left|\mathbf{z}\right| \ \text{where} \ L \mathbf{z} = \left(\mathbf{x} - \bm{\mu}\right), \ \sigma = L L^{T}.
    \end{equation}
    \noindent This last step represents the Cholesky decomposition. This calculation can fail if $\sigma$ is not positive definite.
    \item[3.] If the second method also fails, we instead calculate the pseudo-Mahalanobis distance by using the pseudo-inverse of the covariance matrix:
    \begin{equation}
    d_{\mathrm{pM}} = \sqrt{\left(\mathbf{x} - \bm{\mu}\right)^{T} \sigma^{+} \left(\mathbf{x} - \bm{\mu}\right)},
    \end{equation}
    \noindent where the "$+$" subscript represents the pseudo-inverse matrix operation. This method is guaranteed to work since every matrix has a unique pseudo-inverse.
\end{enumerate}

The fourth measurement method is the permutation importance \citep{Zien2009,Konig2021}, which is also native to \texttt{sklearn}. With permutation importance, we randomly permute each feature to effectively eliminate its correlation with candidacy. For each feature, we then retrain a model with the original dataset but with that feature permuted. We then investigate the $F_{1}$ scores of these models, where a low decrease in $F_{1}$ indicates low importance.

\section{Results} \label{sec:results}

\subsection{Hyperparameter Tuning Results} \label{sec:hyperparameter_tuning_results}

\begin{deluxetable*}{lcccccc}[!ht]
\tablecaption{\label{tab:feature_importance_rankings}Top 5 feature importance rankings via different selection methods. Forward selection recursively removes the least important feature, and vice versa.}
\tablehead{
\colhead{\multirow{2}{*}{Rank}} & \colhead{\multirow{2}{*}{Select K-Best}} & \colhead{\multirow{2}{*}{RFE Forward}} & \colhead{\multirow{2}{*}{RFE Backward}} & \colhead{Mahalanobis} & \colhead{Mahalanobis} & \colhead{Permutation} \\
{ } & { } & { } & { } & \colhead{Forward} & \colhead{Backward} & \colhead{Importance}}
\startdata
1.  & Metallicity      & Metallicity  & Halo Mass    & Metallicity    & Temperature    & Avg. $dM/dz$ \\
2.  & Stellar Mass     & Halo Mass    & Metallicity  & Halo Mass      & Halo Mass      & Halo Mass \\
3.  & LW Flux          & LW Flux      & LW Flux      & LW Flux        & Rad. Vel. Sign & Avg. $dM/dt$ \\
4.  & Density          & Avg. $dM/dt$ & Stellar Mass & RMS Vel.       & Tan. Vel.      & Temperature \\
5.  & Radial Mass Flux & Tan. Vel.    & Temperature  & Rad. Vel. Sign & Avg. $dM/dt$   & $\mathrm{H}_{2}$ Fraction \\
\enddata
\end{deluxetable*}

\begin{figure*}[!ht]
\centering
\includegraphics[width=0.95\textwidth]{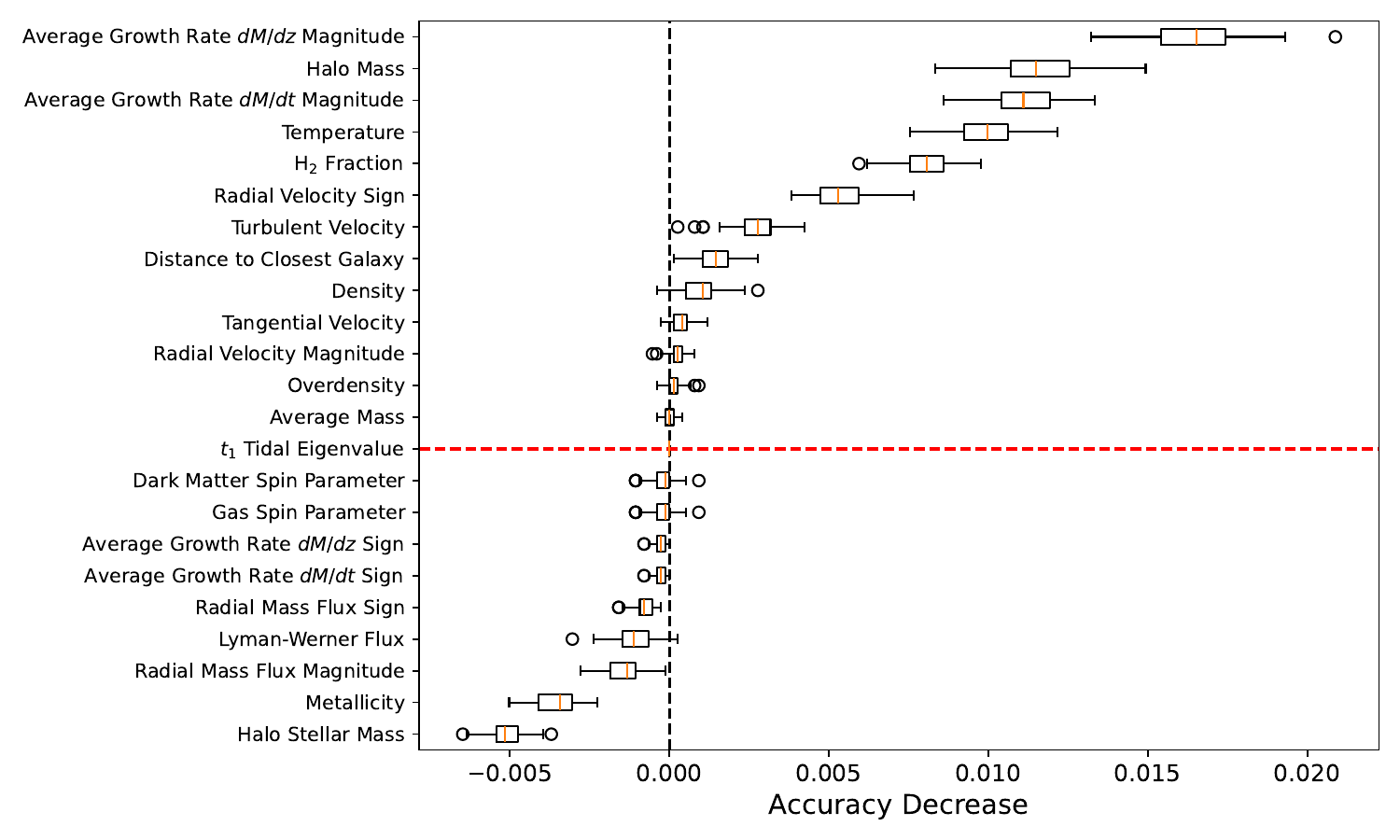}
\caption{\label{fig:permutation_importance}Permutation importance rankings for each feature. The black dashed line represents no decrease in accuracy, and all features at or below the red line show an increase in performance accuracy when permuted. Increases in performance accuracy are likely due to correlated variables \citep{Mone2025} or due to effects from hyperparameter choice and the distributions of candidates vs. non-candidates in phase space (see Section~\ref{sec:model_performance}).}
\end{figure*}

The hyperparameter tuning results for the model with all features are shown in Figure~\ref{fig:F1s_all_features}. All kernels tend to exhibit strong performance at high $C$ and intermediate $w$ that gradually decreases towards the corner of phase space corresponding to low $C$ and high $w$, but with a sharp cutoff approaching the region of phase space defined by low $C$ and low $w$. Models with linear kernels can perform better at intermediate values of both $C$ and $w$ above this cutoff. Models likely have such poor performance at low $w$ due to the unbalanced nature of our dataset, and we can begin to probe lower values of $w$ with acceptable performance by further increasing $C$ since this more heavily penalizes incorrect misclassifications; conversely, models at the other ends of the spectrum likely have such high $w$ values that they focus too heavily on classifying points as candidates without sufficient penalty due to the low $C$ value. Lower $C$ values also increase the size of the margins and the number of support vectors used to calculate the decision boundary. This may be an issue for our dataset since it is heavily unbalanced, meaning that larger margins could lead to more and more non-candidates taken as support vectors on both sides of the decision boundary, which could lead to poorer boundaries as it becomes increasingly difficult to determine which side should correspond to candidates rather than non-candidates. The $F_{1}$ scores for the sigmoid kernel are consistently low, which is expected because sigmoid typically provides a good fit only in very specialized situations, which is likely not the case with our dataset. The model with the maximum $F_{1}$ score uses a 5th-order polynomial kernel with $C = 10^{3}$ and $w = 10^{1.25}$,  with a performance of $F_{1} = 0.325$ determined by the \texttt{GridSearchCV} function in \texttt{sklearn}.

\subsection{Feature Selection} \label{sec:feature_selection}

Using this model, we employed the feature importance measurement methods described in Section~\ref{sec:feature_importance}. We show the top 5 ranked features for each of our feature importance methods in Table~\ref{tab:feature_importance_rankings}; the full table is available in Appendix~\ref{apx:table}. Features like halo mass, metallicity, and LW flux often ranked among the top features for the methods we tested. Features like stellar mass and the average halo growth rates have more variable rankings depending on the method used, despite the fact that stellar mass was one of the variables used to define a DCBH candidate \citep{Mone2025}. This variation in rankings highlights the necessity of testing different feature importance methods.

We also show boxplots of the permutation importance rankings for each feature in Figure~\ref{fig:permutation_importance}. For this measurement, the model is retrained 100 times with one feature randomly permuted (thereby removing any correlation with candidacy), and the decrease in accuracy relative to the original model is calculated. A positive decrease highlights that permuting the feature removes valuable information that the model uses to produce a good fit, thereby corresponding to a higher feature importance.

\begin{figure}[!ht]
\centering
\includegraphics[width=0.99\linewidth]{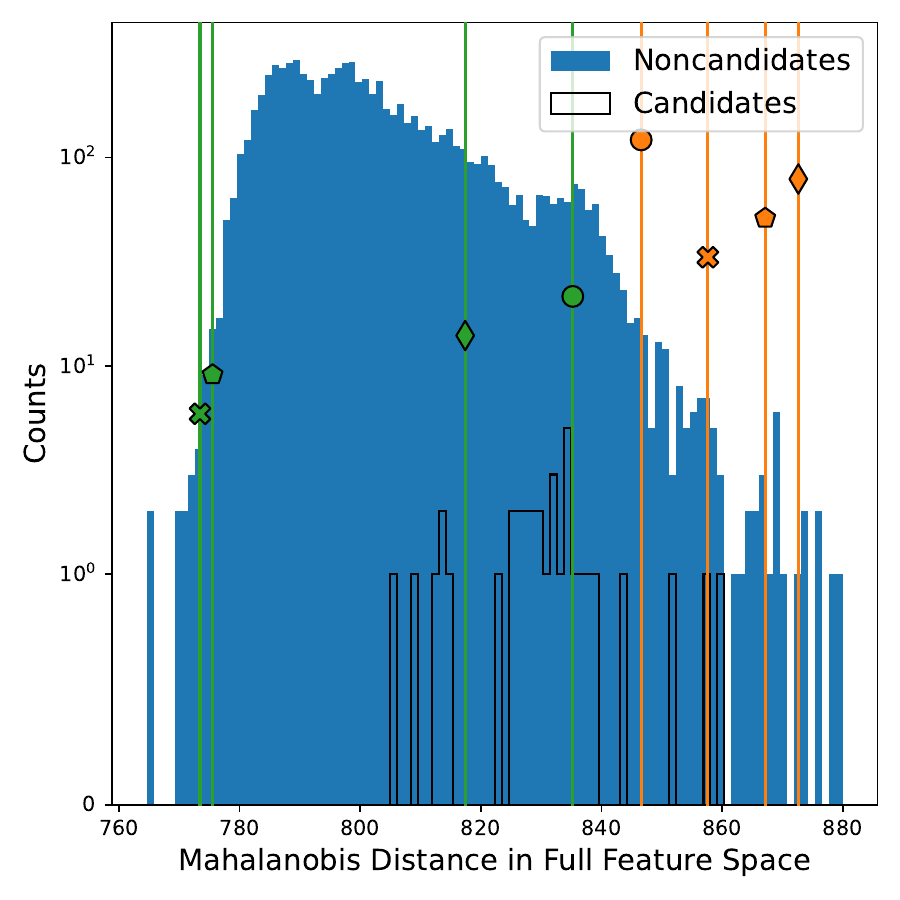}
\caption{\label{fig:Mahalanobis_distribution}Distribution of Mahalanobis distances in the full feature space relative to the mean of the candidate set for both non-candidates (filled blue) and candidates (black). Halos marked with solid lines and markers were non-candidates misclassified by at least one model and were chosen for further inspection, described in Section~\ref{sec:noncandidate_properties}. The markers are vertically offset from each other to prevent overlap.}
\end{figure}

\begin{deluxetable}{c|cc}[!ht]
\tablecaption{\label{tab:Mahalanobis_summary_statistics}Summary statistics for the Mahalanobis distance distributions of candidates and non-candidates.}
\tablehead{\colhead{Statistic} & \colhead{Candidates} & \colhead{Non-candidates}}
\startdata
Mean & 830.31 & 803.12 \\
Median & 830.90 & 799.31 \\
Standard Deviation & 11.85 & 17.03 \\
Minimum & 805.69 & 764.59 \\
Maximum & 859.88 & 880.00 \\
$\pm 1\sigma$ & 837.79, 817.78 & 820.90, 786.71 \\
$\pm 2\sigma$ & 857.86, 808.30 & 842.10, 779.89 \\
$\pm 3\sigma$ & 859.76, 805.84 & 869.19, 773.49
\enddata
\end{deluxetable}

\begin{deluxetable*}{c|cccccc}[htb!]
\tablecaption{\label{tab:feature_subsets}Inclusion of SVM features in different feature subsets.}
\tablehead{\colhead{Feature} & \colhead{\texttt{dm\_main}} & \colhead{\texttt{dm\_full}} & \colhead{\texttt{gas\_main}} & \colhead{\texttt{gas\_full}} & \colhead{\texttt{star\_main}} & \colhead{\texttt{star\_full}}}
\startdata
Halo Mass                 & Y   & Y   & Y   & Y   & Y   & Y \\
Spin Param. DM            & Y   & Y   & Y   & Y   & Y   & Y \\
Avg. $dM/dz$              & $-$ & Y   & $-$ & Y   & $-$ & Y \\
Avg. $dM/dz$ Sign         & $-$ & Y   & $-$ & Y   & $-$ & Y \\
Temperature               & $-$ & $-$ & Y   & Y   & Y   & Y \\
Rad. Vel.                 & $-$ & $-$ & Y   & Y   & Y   & Y \\
Rad. Vel. Sign            & $-$ & $-$ & Y   & Y   & Y   & Y \\
RMS Vel.                  & $-$ & $-$ & Y   & Y   & Y   & Y \\
Radial Mass Flux          & $-$ & $-$ & Y   & Y   & Y   & Y \\
Radial Mass Flux Sign     & $-$ & $-$ & Y   & Y   & Y   & Y \\
Spin Param. Gas           & $-$ & $-$ & Y   & Y   & Y   & Y \\
Density                   & $-$ & $-$ & $-$ & Y   & $-$ & Y \\
Metallicity               & $-$ & $-$ & $-$ & $-$ & Y   & Y \\
LW Flux                   & $-$ & $-$ & $-$ & $-$ & Y   & Y \\
Stellar Mass              & $-$ & $-$ & $-$ & $-$ & Y   & Y \\
$\mathrm{H}_{2}$ Fraction & $-$ & $-$ & $-$ & $-$ & $-$ & Y \\
\hline
Tan. Vel.                 & $-$ & $-$ & $-$ & $-$ & $-$ & $-$ \\
Overdensity               & $-$ & $-$ & $-$ & $-$ & $-$ & $-$ \\
Closest Galaxy            & $-$ & $-$ & $-$ & $-$ & $-$ & $-$ \\
Avg. Halo Mass            & $-$ & $-$ & $-$ & $-$ & $-$ & $-$ \\
Avg. $dM/dt$              & $-$ & $-$ & $-$ & $-$ & $-$ & $-$ \\
Avg. $dM/dt$ Sign         & $-$ & $-$ & $-$ & $-$ & $-$ & $-$ \\
$t_{1}$ Eigenvalue        & $-$ & $-$ & $-$ & $-$ & $-$ & $-$ \\
\enddata
\end{deluxetable*}

Figure~\ref{fig:Mahalanobis_distribution} shows the distribution of Mahalanobis distances for each halo relative to the candidate set in the full feature space. We also present summary statistics of these distributions in Table~\ref{tab:Mahalanobis_summary_statistics}. Surprisingly, there are many non-candidates with a smaller Mahalanobis distance relative to the mean of the candidate set. By construction, we expect that most of the candidates will have small separations for certain features, such as stellar mass (since $M_{\star} = 0$ by definition), halo mass, and metallicity. However, the halo mass cut on our sample ensures that all halos should have relatively small separations in mass, and many non-candidates also have zero stellar mass. This suggests that the candidates have a large spread for certain features that lead to larger Mahalanobis distances. These would be features that have larger value ranges (for example, metallicity spans $\sim\!20$ orders of magnitude, leading to a range of $20$; contrast this with halo mass, with a range of $<0.5$ due to the halo mass cut). Note that this is regarding the spread in the differences between the halo values and the mean values for the candidate set, not the spread in the Mahalanobis distances; the candidates have a lower standard deviation in their Mahalanobis distances than the non-candidates, but the mean and median Mahalanobis distances are larger for the candidates than the non-candidates.

Since the goal of this work is to develop one or more SVMs that can be used in cosmological simulations, we created feature subsets that would be relevant for different types of simulations or semi-analytic models: those that model only dark matter (\texttt{dm}); those that model dark matter and gas (\texttt{gas}); and those that model dark matter, gas, and stars (\texttt{star}). Then, with the results from the feature importance rankings, we developed for each of these simulation types a subset of primary features (\texttt{main}), and a subset that included extra features (\texttt{full}) that were of lower importance, were already correlated with one of the primary features, or would be more difficult to calculate on-the-fly within a cosmological simulation. Some of our original features are also excluded from all subsets for these reasons. The features included in each of these subsets are shown in Table~\ref{tab:feature_subsets}.

The halo mass and DM spin parameter are the primary features available for a DM-only simulation. The average growth rate ${dM}/{dz}$ may also be useful if merger trees for the DM halos are available, which is why we relegated it to a secondary variable. Several important features are added when gas is in play: the temperature, radial velocity, turbulent velocity, radial mass flux, and gas spin parameter. The density is also available here, but given that we expect this to be correlated with most of the other features, we relegated it to a backup variable. Additionally, though high density indicates the formation of some central structure, it is not necessarily conducive for SMS formation; other thermodynamic variables are more relevant for distinguishing the formation of a DCBH from some other central structure like a nuclear star cluster. When star formation is turned on, we add the metallicity, LW flux, and halo stellar mass features. The $\mathrm{H}_{2}$ fraction is also available here, but since this has been shown to be correlated with the metallicity and LW flux \citep{Mone2025}, we relegated it to a backup variable. This leaves the following features excluded from all subsets: tangential velocity, overdensity, distance to the closest galaxy, average halo mass, average growth rate ${dM}/{dt}$, and $t_{1}$ tidal eigenvalue. These variables have been excluded for at least one of the following reasons: low overall importance, degeneracy or high correlation with another feature, or computational complexity. For example, the $t_{1}$ tidal eigenvalue, the overdensity, the average halo mass, and the average growth rate ${dM}/{dt}$ are all expensive and burdensome to compute, the latter two of which require halo merger histories. The average growth rate ${dM}/{dt}$ is degenerate with ${dM}/{dz}$, and the LW flux serves as a proxy for the distance to the closest galaxy.

\subsection{Model Performance} \label{sec:model_performance}

\begin{figure}[!ht]
\centering
\includegraphics[width=0.99\linewidth]{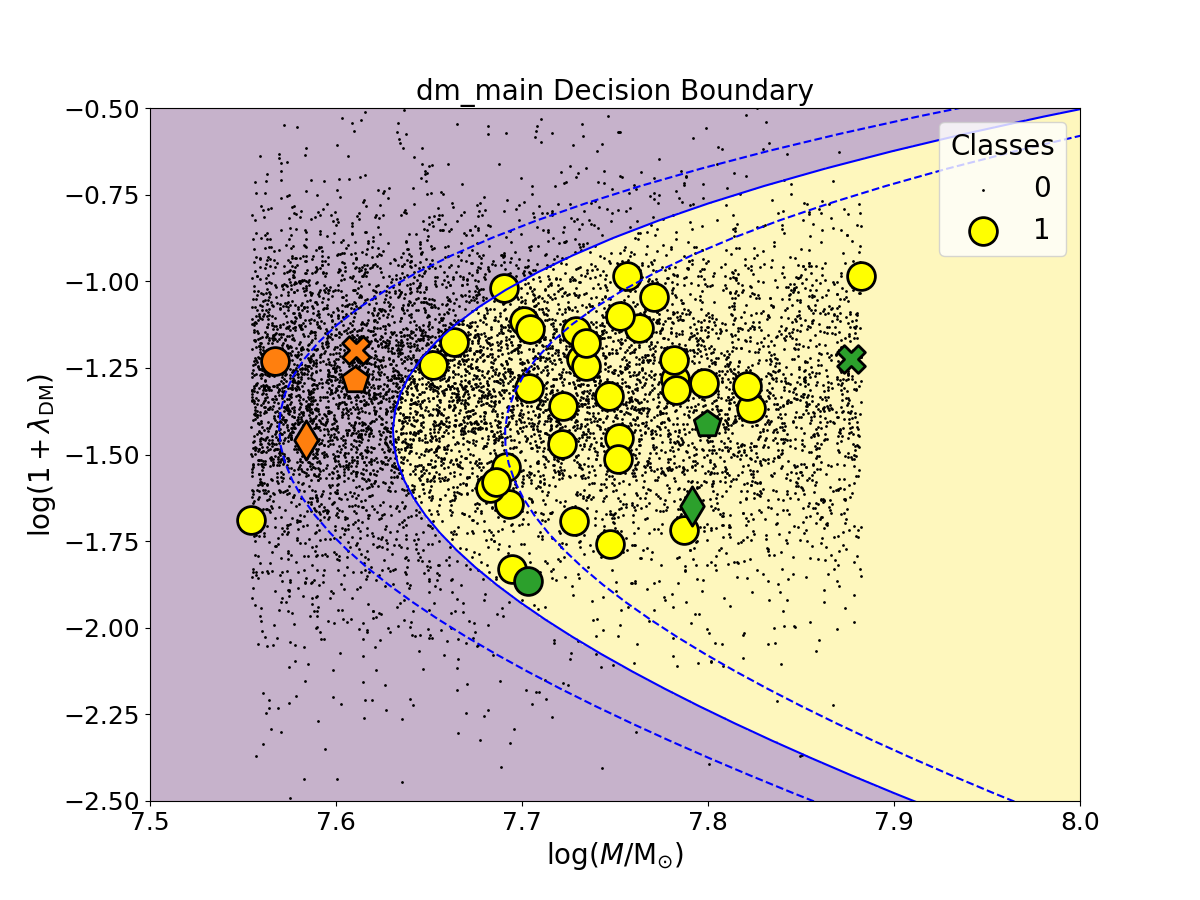}
\caption{\label{fig:decision_boundaries_dm_main}Decision boundary for the best \texttt{dm\_main} model, with a 3rd-order polynomial kernel, $C = 10^{4}$, and $w = 10^{2.5}$. The yellow circles and black points correspond to candidates and non-candidates, respectively, while the orange and green markers identify the same halos described in Figure~\ref{fig:Mahalanobis_distribution}. The yellow and purple regions show the regions of phase space that the SVM determines correspond to candidates and non-candidates, respectively, while the solid blue line shows the decision boundary and the dashed blue lines show the margins on either side of the boundary. This model uses the largest possible value for $C$, corresponding to the smallest margins and few support vectors. The SVM correctly finds the region of phase space occupied by all but 1 of the candidates.}
\end{figure}

With these subsets in place, we re-tuned the hyperparameters for each subset. The best models for each subset are shown in Table~\ref{tab:subset_performance}. Most subsets prefer polynomial kernels, highlighting a need for nonlinearity, and large $C$ and intermediate $w$ values for the same reasons as discussed in Section~\ref{sec:feature_selection}. It is also worth noting that $w$ tends to decrease as more features are added, likely because additional features lead to better separation in phase space and therefore require less weighting of the candidate data points to yield a reasonable fit.

\begin{deluxetable*}{c|cDDDD}[!ht]
\tablecaption{\label{tab:subset_performance}Model with the best \texttt{GridSearchCV} performance for each feature subset. $F_{1}$ values are rounded to three significant digits.}
\tablehead{\colhead{Feature Subset} & \colhead{Kernel} & \multicolumn{2}{c}{$\log_{10}\!\left(C\right)$} & \multicolumn{2}{c}{$\log_{10}\!\left(w\right)$} & \multicolumn{2}{c}{\texttt{GridSearchCV} $F_{1}$} & \multicolumn{2}{c}{\texttt{SVC} $F_{1}$}}
\decimals
\startdata
\texttt{dm\_main}   & Polynomial, $d=3$ & 4.00 & 2.50 & 0.0129 & 0.0197 \\
\texttt{dm\_full}   & RBF               & 4.00 & 2.00 & 0.0296 & 0.0 \\
\texttt{gas\_main}  & Polynomial, $d=5$ & 4.00 & 1.50 & 0.0867 & 0.0 \\
\texttt{gas\_full}  & Polynomial, $d=5$ & 4.00 & 1.50 & 0.0672 & 0.333 \\
\texttt{star\_main} & Polynomial, $d=3$ & 3.75 & 0.75 & 0.372  & 0.316 \\
\texttt{star\_full} & Polynomial, $d=5$ & 3.50 & 1.25 & 0.368  & 0.240 \\
\enddata
\end{deluxetable*}

At this stage, we also show the decision boundaries for the best \texttt{dm\_main} model in Figure~\ref{fig:decision_boundaries_dm_main}. Of the six models listed in Table~\ref{tab:subset_performance}, this is the only one we can show the decision boundaries for because it is the only model with a two-dimensional phase space. We find that the SVM does correctly identify the region of phase space populated by the DCBH candidate halos, but this region is also saturated with non-candidates, leading to the low $F_{1}$ score. In short, this model does not have sufficient information to properly distinguish between the candidates and non-candidates. We also note that halos shown in green markers lie in the candidate region and those shown in orange markers lie in the non-candidate region, which corresponds to their Mahalanobis distances (see Figure~\ref{fig:Mahalanobis_distribution}).

As the \texttt{dm\_main} model is the only model with only two features, it is not tractable for us to attempt to visualize the decision boundaries of models in higher-dimensional spaces. The complexity of these decision boundaries and the loss of critical information by projecting the surfaces down to lower dimensions would not yield sensible results. We instead trained models in specific 2D feature subspaces, focusing on features critical to DCBH formation. We inspect all six pairs of the following features: metallicity, LW flux, radial mass flux, and stellar mass. By focusing on these features, we expect to get better performance in a 2D space than the \texttt{dm\_main} model.

The best models for each of these 2D subspaces are presented in Table~\ref{tab:subspace_performance}. Most models prefer polynomial kernels and $w \approx 10^{1.5}$. All of these models have a performance of $F_{1} \approx 0.1$, which is significantly improved over the \texttt{dm\_main} model with $F_{1} = 0.0129$.

\begin{deluxetable*}{cc|cDDDD}[!ht]
\tablecaption{\label{tab:subspace_performance}Model with the best \texttt{GridSearchCV} performance for each 2D feature subspace.}
\tablehead{\colhead{Feature 1} & \colhead{Feature 2} & \colhead{Kernel} & \multicolumn{2}{c}{$\log_{10}\!\left(C\right)$} & \multicolumn{2}{c}{$\log_{10}\!\left(w\right)$} & \multicolumn{2}{c}{\texttt{GridSearchCV} $F_{1}$} & \multicolumn{2}{c}{\texttt{SVC} $F_{1}$}}
\decimals
\startdata
Metallicity      & LW Flux          & Polynomial, $d=3$ &  3.25 & 1.25 & 0.133  & 0.0930 \\
Metallicity      & Radial Mass Flux & Polynomial, $d=5$ &  4.00 & 1.25 & 0.0968 & 0.235 \\
Metallicity      & Stellar Mass     & Linear            &  4.00 & 1.75 & 0.124  & 0.156 \\
LW Flux          & Radial Mass Flux & Polynomial, $d=2$ &  1.50 & 1.50 & 0.120  & 0.174 \\
LW Flux          & Stellar Mass     & Polynomial, $d=4$ & -0.25 & 1.50 & 0.0914 & 0.0563 \\
Radial Mass Flux & Stellar Mass     & Polynomial, $d=5$ &  3.25 & 1.25 & 0.132  & 0.182 \\
\enddata
\end{deluxetable*}

\begin{figure*}[!ht]
\centering
\includegraphics[width=0.95\textwidth]{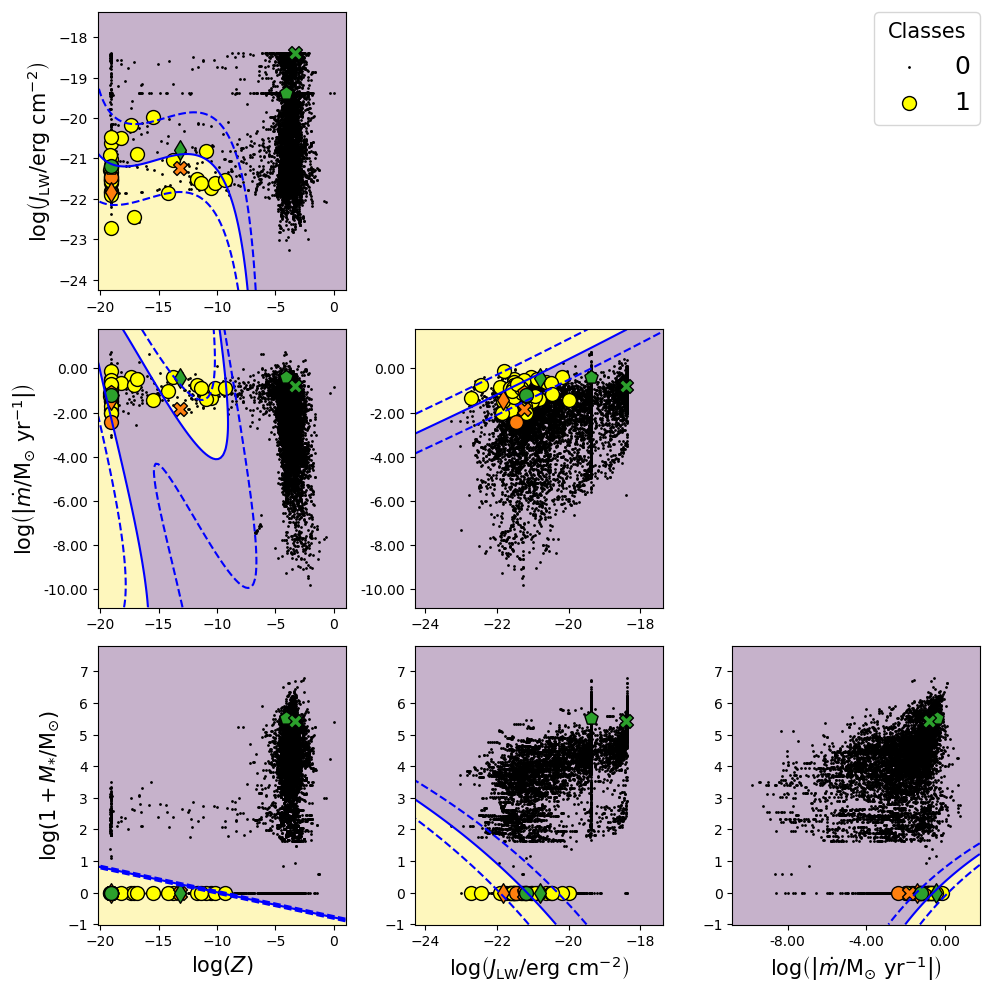}
\caption{\label{fig:decision_bounadries_subspaces}Decision boundaries for the six subspaces of feature pairs. Points and lines are the same as described in Figure~\ref{fig:decision_boundaries_dm_main}.}
\end{figure*}

\begin{figure*}[!ht]
\centering
\includegraphics[width=0.95\textwidth]{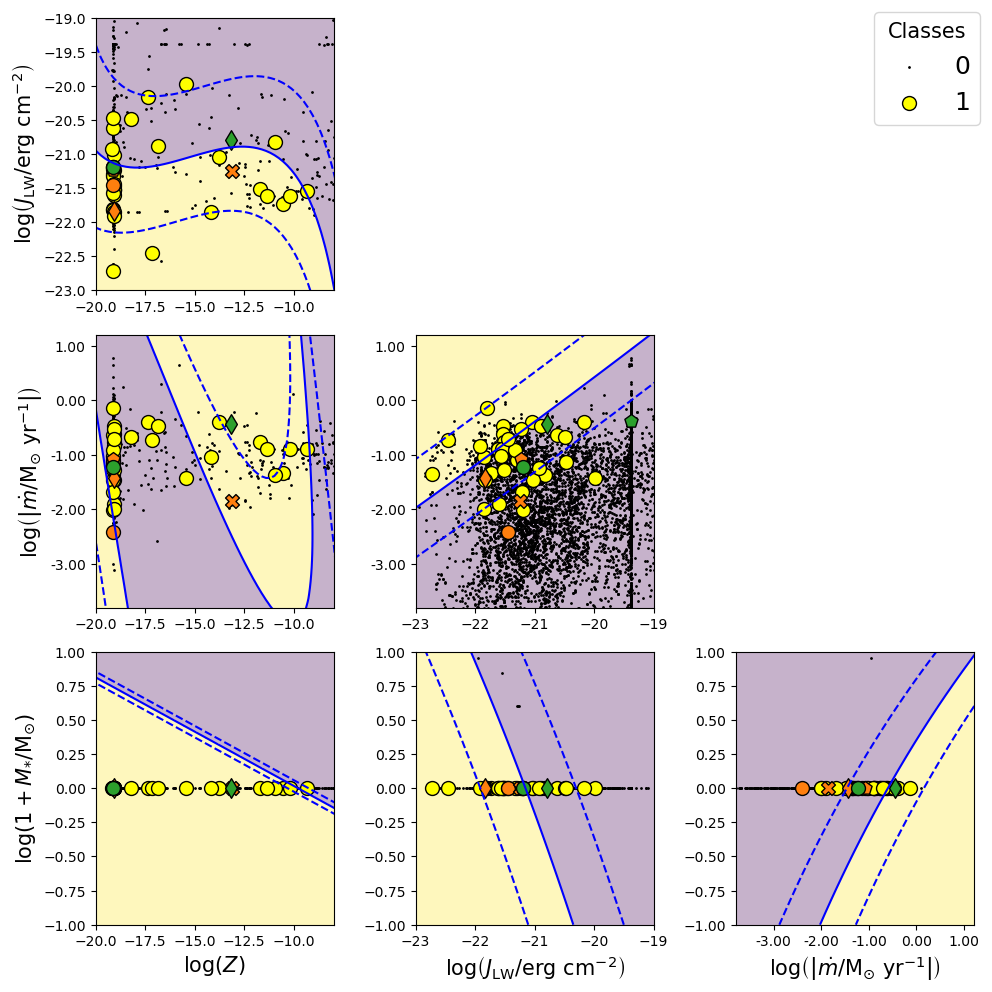}
\caption{\label{fig:decision_bounadries_zoom_subspaces}Same as Figure~\ref{fig:decision_bounadries_subspaces} zoomed in on the DCBH candidate region of each phase space.}
\end{figure*}

We also show the corresponding decision boundaries for these models in Figure~\ref{fig:decision_bounadries_subspaces}, with zoom-ins around the DCBH candidate regions of phase space shown in Figure~\ref{fig:decision_bounadries_zoom_subspaces}. These phase spaces exhibit better separation of candidates and non-candidates, which improves overall performance. However, most of these models are unable to fully capture the DCBH region due to penalties from misclassifying non-candidates in the same regions. This behavior is an artifact of both the hyperparameters and how the SVM minimizes error. In the \texttt{dm\_main} model where the candidates and non-candidates are not well-separated in phase space, the SVM minimizes error by attempting to capture as many candidates as possible. In these subspace models where the separation in phase space is clearer, the highest reduction in error comes from excluding non-candidates, even at the cost of also excluding candidates.

For example, consider the metallicity--LW flux ($Z-J_{\text{LW}}$) subspace in the upper left panel of Figure~\ref{fig:decision_bounadries_subspaces}. The decision boundary here is smooth, but contrary to expectations, the SVM identifies the candidate regions as those with lower LW fluxes. This does a decent job of capturing where the candidates sit in phase space, but fails to capture the candidates with $J_{\text{LW}} \ge 10^{-21} \, \mathrm{{erg} \ {cm}^{-2}}$ because it gets a higher score by excluding the non-candidates in this region. This model also highlights a sharp cutoff near $Z \sim 10^{-8}$, indicating that it learns that all of the candidates are metal-free.

As a counterexample, consider the metallicity--radial mass flux ($Z-\dot{m}$) subspace in the middle left panel. The decision boundary here is more exotic despite having a polynomial kernel like in the $Z-J_{\text{LW}}$ subspace (albeit with a higher polynomial order of $5$ rather than $3$). However, the underlying behavior is still the same. This SVM captures the region of phase space where many of the candidates sit (high $\dot{m}$ and intermediate $Z$), while ignoring the region with a higher concentration of non-candidates closer to $Z \sim 10^{-20}$, which corresponds to halos that have had zero metal enrichment from the initial simulation value. It is worth noting that the wide margins of this SVM do fully capture the candidate population without accepting regions of phase space heavily dominated by non-candidates.

\subsection{Properties of Non-candidate Halos} \label{sec:noncandidate_properties}

To better understand why some non-candidates were misclassified as candidates, we identified the non-candidates misclassified by any of the twelve models described in Section~\ref{sec:model_performance} and sorted them into four categories: those that were cooling but not collapsing, those that were cooling and collapsing, those that were experiencing stellar radiative feedback, and those that were experiencing supernova feedback. We then selected the non-candidates from each category that had the largest and smallest Mahalanobis distances (see Figure~\ref{fig:Mahalanobis_distribution}). The properties of these halos are shown in Table~\ref{tab:misclassified_halos_largest} (largest distances) and Table~\ref{tab:misclassified_halos_smallest} (smallest distances). We notice that, of the non-candidates selected for investigation, those with larger Mahalanobis distances tend to be misclassified by the 2D subspace models, and often by models that consider LW and\slash or those that do not consider the radial mass flux. Conversely, those with the smaller Mahalanobis distances tend to be misclassified by one of our original feature subsets (particularly the \texttt{dm\_main} model), and the non-candidates that are not experiencing feedback are misclassified by several models, highlighting the similarities between these halos and the candidates and the subsequent difficulty in distinguishing these non-candidates from the candidates.

\begin{deluxetable*}{c|cDDcD}[!ht]
\tablecaption{\label{tab:misclassified_halos_largest}Halos misclassified by at least one model with the largest Mahalanobis distance from the candidate set in the full feature space for each category. Also shown are the model(s) in which the halo was misclassified and the corresponding Mahalanobis distances in each model's feature space. These halos are identified with orange markers: a circle, diamond, pentagon, and X, respectively.}
\tablehead{\multicolumn{1}{c|}{\multirow{2}{*}{Category}} & \colhead{Halo Mass} & \multicolumn{2}{c}{\multirow{2}{*}{Redshift}} & \multicolumn{2}{c}{Mahalanobis} & \colhead{\multirow{2}{*}{Model(s)}} & \multicolumn{2}{c}{Model} \\
{ } & \colhead{$\left[\mathrm{M}_{\odot}\right]$} & { } & { } & \multicolumn{2}{c}{Distance} & { } & \multicolumn{2}{c}{Distances}}
\decimals
\startdata
\multirow{4}{*}{No feedback, not cooling} & \multirow{4}{*}{$3.69 \times 10^{7}$} & \multirow{4}{*}{17.0} & \multirow{4}{*}{846.582} & \texttt{metallicity\_LW} & 2.179 \\
{ } & { } & { } & { } & \texttt{metallicity\_radial\_mass\_flux} & 51.497 \\
{ } & { } & { } & { } & \texttt{LW\_stellar\_mass} & 2.401 \\
{ } & { } & { } & { } & \texttt{LW\_stellar\_mass} & 19.304 \\
\hline
\multirow{3}{*}{No feedback, cooling    } & \multirow{3}{*}{$3.83 \times 10^{7}$} & \multirow{3}{*}{16.8} & \multirow{3}{*}{872.638} & \texttt{metallicity\_LW} & 1.925 \\
{ } & { } & { } & { } & \texttt{metallicity\_stellar\_mass} & 2.404 \\
{ } & { } & { } & { } & \texttt{LW\_stellar\_mass} & 18.959 \\
\hline
\multirow{3}{*}{Stellar feedback        } & \multirow{3}{*}{$4.08 \times 10^{7}$} & \multirow{3}{*}{15.5} & \multirow{3}{*}{867.144} & \texttt{metallicity\_LW} & 2.340 \\
{ } & { } & { } & { } & \texttt{metallicity\_stellar\_mass} & 2.400 \\
{ } & { } & { } & { } & \texttt{LW\_stellar\_mass} & 19.506 \\
\hline
\multirow{4}{*}{Supernova feedback      } & \multirow{4}{*}{$4.08 \times 10^{7}$} & \multirow{4}{*}{16.3} & \multirow{4}{*}{857.641} & \texttt{metallicity\_LW} & 7.552 \\
{ } & { } & { } & { } & \texttt{metallicity\_radial\_mass\_flux} & 44.535 \\
{ } & { } & { } & { } & \texttt{metallicity\_stellar\_mass} & 3.354 \\
{ } & { } & { } & { } & \texttt{LW\_stellar\_mass} & 19.493 \\
\enddata
\end{deluxetable*}

\begin{figure*}[!ht]
\centering
\includegraphics[width=0.9\textwidth]{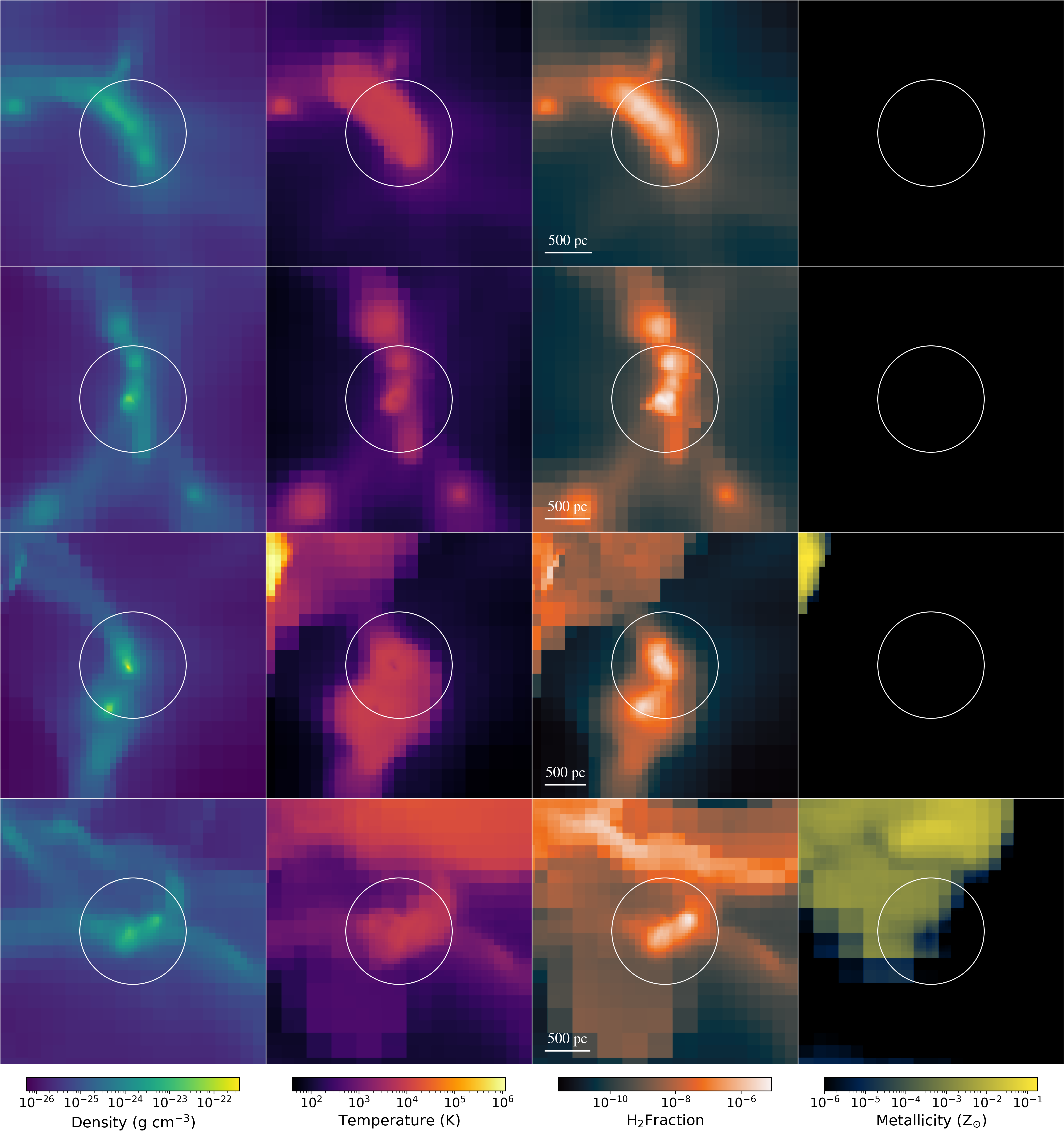}
\caption{\label{fig:projections_z_Mahalanobis_large}Density-weighted projections of misclassified halos with large Mahalanobis distances relative to the candidate set. Rows correspond to the same halos described in Table~\ref{tab:misclassified_halos_largest} (no feedback and not cooling, no feedback and cooling, stellar radiative feedback, and supernova feedback, respectively). The width of each panel and the depth of each projection are five times the halo's virial radius, which is represented by the white circle.}
\end{figure*}

\begin{figure*}[!ht]
\centering
\includegraphics[width=0.9\textwidth]{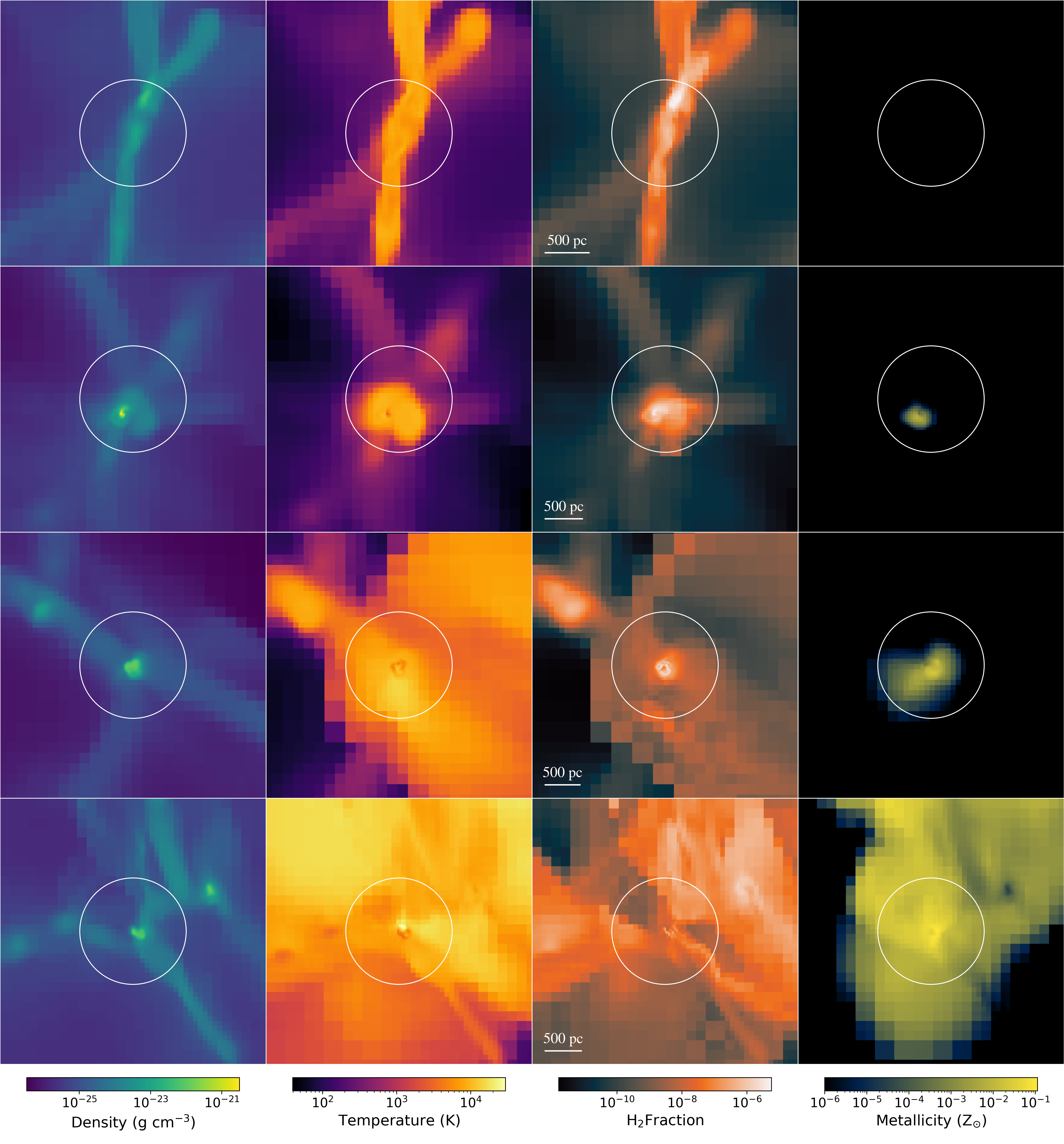}
\caption{\label{fig:projections_z_Mahalanobis_small}Same as Figure~\ref{fig:projections_z_Mahalanobis_large} for misclassified halos with small Mahalanobis distances relative to the candidate set. Rows correspond to the same halos described in Table~\ref{tab:misclassified_halos_largest} (no feedback and not cooling, no feedback and cooling, stellar radiative feedback, and supernova feedback, respectively).}
\end{figure*}

Figure~\ref{fig:projections_z_Mahalanobis_large} and Figure~\ref{fig:projections_z_Mahalanobis_small} show density-weighted projections for quantities of interest for each of these halos. As expected, the halos that are not experiencing feedback are entirely or almost entirely metal-free, and those that are experiencing feedback show more prominent regions that are metal-enriched. Regions of high temperature coincide with the elevated metallicities and with elevated $\mathrm{H}_{2}$ fractions from recent supernovae. However, the regions with the highest $\mathrm{H}_{2}$ fractions are those at the centers of gas-rich halos that are cooling, collapsing, and undisturbed by stellar feedback. The higher densities lead to an increased $\mathrm{H}_{2}$ formation rate and thus colder temperatures. Some of these halos, particularly the ones in the lower rows of Figure~\ref{fig:projections_z_Mahalanobis_small}, show extended regions with higher $\mathrm{H}_{2}$ fractions despite higher temperatures. These are locations of ionization fronts where the partial ionization of hydrogen leads to an increase in free electrons, which catalyze the formation of $\mathrm{H}_{2}$ \citep{Ricotti2001}. Although these regions are likely created by radiative feedback from stars, they are also unlikely to host future star formation since their low densities mean they may not be self-gravitating. In the case of the halo with a larger Mahalanobis distance identified as experiencing stellar feedback (third row of Figure~\ref{fig:projections_z_Mahalanobis_large}), the cool gas in the halo suggests that it is too far away from the neighboring halo to be significantly impacted by ionizing stellar radiation; however, it likely experiencing a strong LW flux from this neighboring halo. There is still $\mathrm{H}_{2}$ present in the center, which may be due to self-shielding \citep{Wolcott-Green2011(a)}.

\section{Discussion} \label{sec:discussion}

\subsection{Implications for DCBHs} \label{sec:dcbh_implications}

Some observations report potential candidates for DCBHs \citep{Agarwal2016(a),Pacucci2016,Smith(Aaron)2016,Kovacs2024,Natarajan2024,Juodzbalis2025(b)}, including recent evidence for the formation of a DCBH at low redshift ($z = 1.14)$ in the $\infty$ Galaxy \citep{vanDokkum2025(a),vanDokkum2025(b)}. Based on our definition of a DCBH candidate, we expect that DCBHs could form in $\sim0.5\%$ of ACL halos in an overdense region of the universe at $z \gtrsim 15$. However, our definition of a candidate does not consider effects such as the incident LW flux and gas inflow rate, both of which could dissociate $\mathrm{H}_{2}$ and allow for a less stringent metallicity threshold in the definition. Therefore, this estimate may reasonably be considered a lower limit for the DCBH occupation fraction for ACL halos in a large-scale overdensity.

\subsection{Implications for Other Seeding Pathways} \label{sec:seeding_implications}

Observations of SMBHs in the early universe—from those with $M_{\bullet} \gtrsim 10^{8} \, \mathrm{M}_{\odot}$ like GNz-11 \citep{Bunker2023,Cameron2023}, UHZ-1 \citep{Goulding2023,Bogdan2024,Natarajan2024}, GHZ9 \citep{Kovacs2024,Napolitano2025(b)}, and others \citep{Agarwal2016(b),Abuter2024,Bosman2024,Furtak2024,Juodzbalis2024,Marshall2025,Silverman2025,Stone2025,Tripodi2025}; to those at the $M_{\bullet} \sim 10^{6} \, \mathrm{M}_{\odot}$ mass scale \citep{Larson2023,Onoue2023,Ananna2024,Maiolino2024(a),Maiolino2024(c),Maiolino2024(d)}—suggest that seed black holes underwent periods of intense (potentially super-Eddington) accretion. This is further supported by reported observations of "overmassive" black holes; that is, black holes with an elevated $M_{\bullet}/M_{\star}$ ratio for their host galaxy relative to the local $M_{\bullet}-M_{\star}$ scaling relation \citep{Agarwal2013,Bogdan2024,Furtak2024,Juodzbalis2024,Maiolino2024(a),Maiolino2024(c),Natarajan2024,Durodola2025,Ji2025(b),Marshall2025,Stone2025,Wu(Yuxuan)2025}. However, these observations do little to constrain the seeding mechanisms for these SMBHs. Our work highlights that a substantial mass inflow to the center of the halo is an important variable for the formation of a DCBH, which agrees with observational indications. However, such a mass inflow is also conducive to other types of black hole seed formation: this could be how large amounts of gas are channeled onto light seeds, fueling the episodes of super-Eddington accretion required to grow them to supermassive scales in such short periods of time; or this could produce a massive star cluster at the center of the halo, which could lead to runaway collapse and the formation of an IMBH.

\subsection{Applications} \label{sec:applications}

One of the most impactful applications of this work is on massive black hole seeding prescriptions in cosmological simulations and semi-analytic models that study the formation and early growth of SMBHs. Due to limitations in resolution and computational resources, many large-volume cosmological simulations use seeding prescriptions that depend only on the halo mass and often seed black hole particles with $M_{\bullet} \gtrsim 10^{5} \, \mathrm{M}_{\odot}$. This has two main drawbacks. First, the seeding prescriptions are likely oversimplified by design for the sake of computational speed and therefore fail to capture the more complex factors inherent in DCBH formation. Secondly, and perhaps more importantly, the starting mass of the black hole particles is often so large that they miss the seeding phase entirely. These simulations study seeds after they have already evolved to $M_{\bullet} \gtrsim 10^{5} \, \mathrm{M}_{\odot}$ even though we stand to gain the most insight on the similarities and differences among seeding mechanisms by studying seeds when they occupy a lower range of masses $\left(\sim\!10^{4} \, \mathrm{M}{_\odot}\right.$ for DCBHs) shortly after birth. We note that there have been efforts to address one of these two points in simulations such as Horizon-AGN \citep{Kaviraj2017}, \texttt{\textsc{Romulus}} \citep{Tremmel2017}, and \textsc{meli{$\odot$}ra} \citep{Cenci2025}, each with more physically motivated criteria for seeding; and \texttt{\textsc{simba}} \citep{Dave2019}, which seeds black holes at $\sim\!10^{4} \, \mathrm{M}_{\odot}$. We note that \textsc{meli{$\odot$}ra} does not have a fixed seeding value, but the provided prescription suggests a minimum of $\approx 8 \times 10^{4} \, \mathrm{M}_{\odot}$, which is comparable to the Horizon-AGN value of $10^{5} \, \mathrm{M}_{\odot}$.

\begin{deluxetable*}{c|cDDcD}[!ht]
\tablecaption{\label{tab:misclassified_halos_smallest}Same as Table~\ref{tab:misclassified_halos_largest} with the smallest Mahalanobis distance from the candidate set in the full feature space for each category. These halos are identified with green markers.}
\tablehead{\multicolumn{1}{c|}{\multirow{2}{*}{Category}} & \colhead{Halo Mass} & \multicolumn{2}{c}{\multirow{2}{*}{Redshift}} & \multicolumn{2}{c}{Mahalanobis} & \colhead{\multirow{2}{*}{Model(s)}} & \multicolumn{2}{c}{Model} \\
{ } & \colhead{$\left[\mathrm{M}_{\odot}\right]$} & { } & { } & \multicolumn{2}{c}{Distance} & { } & \multicolumn{2}{c}{Distances}}
\decimals
\startdata
\multirow{7}{*}{No feedback, not cooling} & \multirow{7}{*}{$5.05 \times 10^{7}$} & \multirow{7}{*}{16.6} & \multicolumn{2}{c}{\multirow{7}{*}{835.242}} & \texttt{dm\_main} & 99.294 \\
{ } & { } & { } & { } & \texttt{dm\_full} & 129.56 \\
{ } & { } & { } & { } & \texttt{star\_main} & 436.226 \\
{ } & { } & { } & { } & \texttt{star\_full} & 617.381 \\
{ } & { } & { } & { } & \texttt{metallicity\_LW} & 2.365 \\
{ } & { } & { } & { } & \texttt{metallicity\_stellar\_mass} & 2.399 \\
{ } & { } & { } & { } & \texttt{LW\_stellar\_mass} & 19.541 \\
\hline
\multirow{8}{*}{No feedback, cooling    } & \multirow{8}{*}{$6.19 \times 10^{7}$} & \multirow{8}{*}{16.2} & \multicolumn{2}{c}{\multirow{8}{*}{817.399}} & \texttt{dm\_main} & 96.806 \\
{ } & { } & { } & { } & \texttt{gas\_main} & 148.090 \\
{ } & { } & { } & { } & \texttt{gas\_full} & 363.285 \\
{ } & { } & { } & { } & \texttt{star\_main} & 418.273 \\
{ } & { } & { } & { } & \texttt{star\_full} & 596.120 \\
{ } & { } & { } & { } & \texttt{metallicity\_radial\_mass\_flux} & 43.076 \\
{ } & { } & { } & { } & \texttt{metallicity\_stellar\_mass} & 3.348 \\
{ } & { } & { } & { } & \texttt{radial\_mass\_flux\_stellar\_mass} & 26.915 \\
\hline
\multirow{1}{*}{Stellar feedback        } & \multirow{1}{*}{$6.30 \times 10^{7}$} & \multirow{1}{*}{15.2} & \multicolumn{2}{c}{\multirow{1}{*}{775.510}} & \texttt{dm\_main} & 94.835 \\
\hline
\multirow{1}{*}{Supernova feedback      } & \multirow{1}{*}{$7.53 \times 10^{7}$} & \multirow{1}{*}{17.1} & \multicolumn{2}{c}{\multirow{1}{*}{773.423}} & \texttt{dm\_main} & 92.683 \\
\enddata
\end{deluxetable*}

The SVMs presented in this work can address both of these shortcomings simultaneously. The SVMs allow for more holistic seeding prescriptions that capture additional environmental factors such as halo stellar mass, metallicity, and radial mass flux. Additionally, the SVMs can serve as probabilistic rather than deterministic seeding prescriptions by having the SVMs predict the probability that a particular halo is a candidate given its properties; this functionality is natively supported in \texttt{sklearn.svm.SVC} using the \texttt{predict\_proba()} method. While a deterministic seeding prescription applied to a particular halo would result in the placement of a seed either 0\% of the time or 100\% of the time, a probabilistic seeding prescription would yield values between these two extremes. The flexibility of the probabilistic prescription may further account for complexities in DCBH seeding. Moreover, these prescriptions do not require formal resolution of the nuclear region of the halo, allowing researchers to save on computational resources by seeding halos at lower simulation resolutions. These two characteristics of an SVM-based seeding prescription allow for physically motivated seeding of DCBHs at the appropriate mass scale without requiring extensive computational resources.

\subsection{Caveats} \label{sec:caveats}

Our work is based on a definition of a DCBH candidate as our original simulations lacked the resolution to confirm the nature of the central structures expected to form at the center of the candidate halos. Though our definition is physically motivated, it may not be perfectly accurate and could include, for example, a halo that may produce an IMBH rather than a DCBH. High-resolution resimulations of the candidate halos to address this shortcoming are planned in our future work.

Additionally, our definition of a candidate could be overly simplified, primarily with the metallicity criterion, as some work suggests that it is possible to form SMSs in halos that are metal-poor instead of metal-free \citep[e.g.,][]{Chon2020}. We could relax this condition in the definition, though this may then require additional criteria for features such as the LW flux and\slash or the radial mass flux to suppress elevated levels of fragmentation expected at higher metallicities.

We chose to focus on SVMs as the ML architecture for this work. Other architectures, such as neural networks, could also be applied. However, we focused on SVMs for three reasons. The first is that SVMs are designed to scale well with the dimensionality of the problem. Neural networks are also capable of tackling high-dimensional, nonlinear problems, but the amount of training time required to achieve similar results in such high-dimensional spaces could pose challenges. The second is interpretability. Since SVMs have a simpler functional form than neural networks, SVMs are more readily human-interpretable. The third is overfitting. Our dataset consists of $\sim\!8400$ data points. However, large and complex machine learning architectures benefit from large datasets (often many millions of data points) to achieve high accuracy. When trained on small datasets, they run the risk of overfitting, meaning that they focus too heavily on patterns present in the training dataset that cause them to generalize poorly to other datasets. With a sample size this small, we reduce the risk of overfitting by employing a model with less complexity.

The SVMs produced in this work are trained to identify DCBHs specifically, meaning that they are not suitable for use with other seeding mechanisms; although they could be used to look at the evolved states of IMBHs and light seeds, this again produces the issue of not investigating the seed at or shortly after birth. However, the procedure outlined in this work could be replicated on a different dataset to develop one or more SVMs that address other seeding channels. Alternatively, an SVM-based seeding prescription for DCBHs could be combined with seeding prescriptions for other seeding mechanisms to produce a simulation that captures multiple formation mechanisms. The \texttt{\textsc{BRAHMA}} simulations \citep{Bhowmick2024(b),Bhowmick2024(c),Bhowmick2025(a)} represent some of the latest work addressing simulations that capture multiple seeding mechanisms.

Furthermore, the exact distinction between a DCBH and an IMBH grows unclear at masses near the boundary we have defined. Realistically, seeds should occupy a continuum of masses for each mechanism that overlap near the mass boundaries; that is to say, the mass boundaries are not hard cutoffs. As alluded to above, it is entirely possible that a DCBH candidate halo we have identified collapses to a nuclear star cluster rather than directly into an SMS. However, depending on the dynamics of this cluster, this could produce several massive stars, light seeds, and IMBHs that merge into a final black hole in the DCBH mass range \citep{Sakurai2017,Regan(John)2020(c),Rantala2025(a)}, or it could lead to stellar collisions producing a single SMS that is then expected to collapse into a DCBH \citep{Regan(John)2018,Solar2025}. This primarily depends on the merger timescale relative to the typical stellar lifetime in the cluster—if the merger timescale is larger, then we expect stars to die and leave behind black holes that could merge into a remnant in the DCBH mass range, but if the merger timescale is smaller, then we expect stars to merge into a single SMS before ending their lives, resulting in a DCBH.

\section{Conclusion} \label{sec:conclusion}

We present several SVMs trained on a subset of halos from the \textit{Renaissance} simulations to distinguish between DCBH-hosting candidate halos and non-candidate halos based on the candidacy definition in \citet{Mone2025}. Our main conclusions are as follows:

\begin{enumerate}
    \item[1.] Feature importance analysis and rankings of the physical variables yield a variety of results based on the method used, highlighting the importance of testing multiple methods and interpreting the results with astrophysical intuition. The feature subsets we choose are physically motivated, based on both the feature importance results and the intent of using SVMs as seeding prescriptions for DCBHs in cosmological simulations. Generally speaking, we find that features related to the definition of candidacy (halo mass, metallicity, and stellar mass) tend to be ranked higher, along with features that tend to be more strongly correlated with candidacy such as the LW flux and the radial mass flux.
    \item[2.] The SVMs presented yield less-than-ideal performance due to the imbalance of our dataset, the feature spaces, and the impact of hyperparameters on the fitting process. In some models, the loss of performance is caused by the number of available features and the degree of separation between candidates and non-candidates in that feature space (such as the \texttt{dm\_main} model); in others, the loss of performance is due to the choice of hyperparameters and the impact this has on how the SVM chooses to minimize errors (like in the 2D subspace models). The \texttt{star\_main} and \texttt{star\_full} models have access to all of the features used in the candidacy definition, meaning that in principle, these models should yield a high degree of separation between candidates and non-candidates in phase space (though this is not possible to visualize due to the number of dimensions). Performance shortcomings here could also be attributed to the fact that the other implicit condition in the candidacy definition is that the simulation cell was at the highest refinement level, which is not information that the SVMs have access to. Access to this fourth condition should result in complete phase space separation.
    \item[3.] Non-candidates misclassified by our models show a variety of physical morphologies, from halos that are still growing to halos that are collapsing to those experiencing feedback from stars and supernovae. Those with larger Mahalanobis distances tend to be misclassified by the 2D subspace models, likely because these models lack the features necessary to distinguish between these non-candidates and the candidates. Those with smaller Mahalanobis distances tend to be misclassified by our original feature subsets, which is likely also due to lack of distinguishing features (like in the \texttt{dm\_main} model) or due to true morphological similarity with the candidates. Visualizations of these halos highlight findings that agree with expectations: halos that are not experiencing feedback tend to have clearer filamentary structures and central regions with high densities and temperatures, particularly for those that are collapsing, while halos that are experiencing feedback show regions of hot, diffuse, metal-enriched gas at larger radii.
\end{enumerate}

The SVMs developed in this work and the hyperparameter tuning code are available to the community at \url{https://github.com/bpries17/DCBH_SVM}; these are also available on Zenodo under an open-source MIT license: \dataset[doi:10.5281/zenodo.19828222]{https://doi.org/10.5281/zenodo.19828222}. These SVMs can be leveraged as holistic, probabilistic seeding prescriptions for DCBHs in cosmological simulations, though researchers should be cognizant of their limitations during implementation and interpretation of the simulation results. These SVMs have the potential to shed new light on the formation and early growth of DCBHs and how these attributes may differ from other seeding mechanisms.

\section{Acknowledgments} \label{sec:acknowledgment}

\begin{acknowledgements}
We kindly thank the anonymous referee for their comments on the draft of the publication. B.\ P.\ and J.\ H.\ W.\ acknowledge funding support from NSF grants AST-2108020 and AST-2510197 and NASA grant 80NSSC21K1053. The analysis for this work was performed on the Phoenix cluster within Georgia Tech's Partnership for an Advanced Computing Environment (PACE). The \textit{Renaissance} simulations were performed on Blue Waters operated by the National Center for Supercomputing Applications (NCSA) with PRAC allocation support by the NSF (awards ACI-0832662, ACI-1238993, and ACI-1514580). This research is part of the Blue Waters sustained-petascale computing project, which is supported by the NSF (awards OCI-0725070 and ACI-1238993) and the state of Illinois. Blue Waters is a joint effort of the University of Illinois at Urbana-Champaign and its NCSA. The freely available astrophysical analysis code \texttt{yt} \citep{Turk2011} and plotting library \texttt{matplotlib} \citep{Hunter2007} were used to construct numerous plots within this paper. Computations described in this work were performed using the publicly available \texttt{\textsc{Enzo}} code, which is the product of a collaborative effort of many independent scientists from numerous institutions around the world.
\end{acknowledgements}

\bibliography{refs}{}
\bibliographystyle{aasjournal}

\appendix
\section{Extended Table} \label{apx:table}

Here we list the full version of Table~\ref{tab:feature_importance_rankings}. Again, features that are typically ranked higher across the different methods tested are often features relevant to DCBH formation: halo mass, metallicity, and LW flux. However, other features like the stellar mass, temperature, density, $\mathrm{H}_{2}$ fraction, radial mass flux, and growth rate see less consistent rankings despite their connection to DCBH formation. This is likely a byproduct of correlations between variables. Some variables, such as the spin parameters, the $t_{1}$ tidal eigenvalue, and the velocity fields, typically receive middle to low rankings, highlighting their lack of importance.

\movetabledown=2in
\begin{rotatetable*}
\begin{deluxetable*}{lcccccc}
\tablecaption{\label{tab:feature_importance_rankings_full}Feature importance rankings via different selection methods. Forward selection recursively removes the least important feature, and vice versa.}
\tablehead{\colhead{Rank} & \colhead{Select K-Best} & \colhead{RFE Forward} & \colhead{RFE Backward} & \colhead{Mahalanobis Forward} & \colhead{Mahalanobis Backward} & \colhead{Permutation Importance}}
\startdata
1.  & Metallicity               & Metallicity               & Halo Mass                 & Metallicity               & Temperature               & Avg. $dM/dz$ \\
2.  & Stellar Mass              & Halo Mass                 & Metallicity               & Halo Mass                 & Halo Mass                 & Halo Mass \\
3.  & LW Flux                   & LW Flux                   & LW Flux                   & LW Flux                   & Rad. Vel. Sign            & Avg. $dM/dt$ \\
4.  & Density                   & Avg. $dM/dt$              & Stellar Mass              & RMS Vel.                  & Tan. Vel.                 & Temperature \\
5.  & Radial Mass Flux          & Tan. Vel.                 & Temperature               & Rad. Vel. Sign            & Avg. $dM/dt$              & $\mathrm{H}_{2}$ Fraction \\
6.  & $\mathrm{H}_{2}$ Fraction & Density                   & Density                   & Closest Galaxy            & Avg. Halo Mass             & Rad. Vel. Sign \\
7.  & Temperature               & Avg. Halo Mass            & $\mathrm{H}_{2}$ Fraction & Radial Mass Flux          & Metallicity               & RMS Vel. \\
8.  & Halo Mass                 & Closest Galaxy            & Tan. Vel.                 & Avg. $dM/dz$              & Stellar Mass              & Closest Galaxy \\
9.  & Rad. Vel. Sign            & $t_{1}$ Eigenvalue        & Rad. Vel. Sign            & Density                   & LW Flux                   & Density \\
10. & Radial Mass Flux Sign     & Spin Param. DM            & Rad. Vel.                 & Temperature               & Density                   & Tan. Vel. \\
11. & Tan. Vel.                 & Avg. $dM/dz$              & Spin Param. Gas           & Tan. Vel.                 & Radial Mass Flux          & Rad. Vel. \\
12. & RMS Vel.                  & Radial Mass Flux          & Spin Param. DM            & Rad. Vel.                 & $\mathrm{H}_{2}$ Fraction & Overdensity \\
13. & Avg. Halo Mass            & Avg. $dM/dz$ Sign         & RMS Vel.                  & $\mathrm{H}_{2}$ Fraction & Radial Mass Flux Sign     & Avg. Halo Mass \\
14. & Avg. $dM/dt$ Sign$^{a}$   & Avg. $dM/dt$ Sign         & Closest Galaxy            & Avg. Halo Mass            & RMS Vel.                  & $t_{1}$ Eigenvalue \\
15. & Avg. $dM/dz$ Sign$^{a}$   & Spin Param. Gas           & Overdensity               & Avg. $dM/dz$ Sign         & Closest Galaxy            & Spin Param. DM \\
16. & Closest Galaxy            & Rad. Vel. Sign            & Radial Mass Flux Sign     & Spin Param. DM            & Spin Param. Gas           & Spin Param. Gas \\
17. & $t_{1}$ Eigenvalue        & $\mathrm{H}_{2}$ Fraction & Radial Mass Flux          & Spin Param. Gas           & Avg. $dM/dt$ Sign         & Avg. $dM/dz$ Sign \\
18. & Avg. $dM/dt$              & Stellar Mass              & Avg. Mass                 & Avg. $dM/dt$ Sign         & Avg. $dM/dz$ Sign         & Avg. $dM/dt$ Sign \\
19. & Overdensity               & Overdensity               & Avg. $dM/dt$ Sign         & Avg. $dM/dt$              & $t_{1}$ Eigenvalue        & Radial Mass Flux Sign \\
20. & Avg. $dM/dz$              & Radial Mass Flux Sign     & Avg. $dM/dt$              & $t_{1}$ Eigenvalue        & Overdensity               & LW Flux \\
21. & Rad. Vel.                 & Temperature               & Avg. $dM/dz$ Sign         & Radial Mass Flux Sign     & Avg. $dM/dz$              & Radial Mass Flux \\
22. & Spin Param. Gas$^{b}$     & Rad. Vel.                 & Avg. $dM/dz$              & Overdensity               & Rad. Vel.                 & Metallicity \\
23. & Spin Param. DM$^{b}$      & RMS Vel.                  & $t_{1}$ Eigenvalue        & Stellar Mass              & Spin Param. DM            & Stellar Mass \\
\enddata
\vspace{1mm}
\noindent $^{a}$ \mbox{The signs of the accretion rates receive the same score using the Select K-Best method, and therefore should be considered to have identical rankings.} \\
\noindent $^{b}$ Same as above for the spin parameters.
\end{deluxetable*}
\end{rotatetable*}

\end{document}